\documentclass[twocolumn,showpacs,preprintnumbers,amsmath,amssymb]{revtex4}

\usepackage{pstricks}
\usepackage{graphicx}
\usepackage{dcolumn}
\usepackage{bm}

\def\beq{\begin{equation}}
\def\eeq{\end{equation}}
\def\bw{\begin{widetext}}
\def\ew{\end{widetext}}

\def\al{\alpha}
\def\bt{\beta}
\def\Ga{\Gamma}
\def\ga{\gamma}

\def\De{\Delta}
\def\ka{\kappa}

\def\te{\theta}

\def\lam{\lambda}

\def\ep{\epsilon}

\def\sq{\sqrt}

\def\l{\left (}
\def\r{\right )}
\def\fr{\frac}
\def\la{\label}
\def\hs{\hspace}
\def\vs{\vspace}

\def\ran{\rangle}
\def\lan{\langle}
\def\ov{\overline}

\def\tm{\times}

\begin{document}


\preprint{BA-06016}

\preprint{CERN-PH-TH/2006-067}


\title{$\te_{13}$, Rare  Processes and Proton Decay in Flipped $SU(5)$
}

\author{Qaisar Shafi$^a$ and
Zurab Tavartkiladze$^{b}$}


\address{{\it $^a$ Bartol Research Institute, Department of Physics and
Astronomy, University of Delaware, Newark, DE 19716, USA\\
$^b$ Physics Department, Theory Division, CERN, CH-1211 Geneva 23, Switzerland}
}

\begin{abstract}
\vspace{0.3cm}

We consider an extended  flipped $SU(5)$ model, supplemented
by a flavor ${\cal U}(1)$ symmetry, which yields bi-large neutrino mixings,
charged fermion mass hierarchies and CKM mixings. The third leptonic
mixing angle $\te_{13}$ turns out to lie close to $0.07$,
and neutrino CP violation can be estimated from the observed baryon asymmetry.
For lepton flavor violating processes we find the branching ratios,
${\rm BR}(\mu \to e\ga )\sim {\rm BR}(\tau \to e \ga )
\sim 10^{-4}\cdot {\rm BR}(\tau \to \mu \ga )
\stackrel{<}{_\sim }5\cdot 10^{-14}$. The proton lifetime
$\tau_{p\to \pi^0 e^{+}}\simeq 10^{34}-10^{36}$~yrs.

\end{abstract}


\maketitle

 Neutrino oscillation experiments \cite{Fukuda:2000np} have provided
strong evidence
for physics beyond the Standard Model and its minimal supersymmetric
extension (MSSM). The two leptonic mixing angles $\theta_{12},
\theta_{23}$ are large, while the third mixing angle
$\theta_{13}\stackrel{<}{_\sim}0.2$.
An unsuppressed value of $\te_{13}$($\stackrel{>}{_\sim }10^{-2}$) would
lead to observable CP violation in the lepton sector.
However, if $\te_{13}$ will turn out to be$\ll 1$, this
may hint to some underlying symmetry.
Since there are not many models \cite{Albrow:2005kw}
which predict $\te_{13}$ by symmetry
arguments, it is desirable to construct models which predict $\te_{13}$.
The task is not easy because knowledge of mixings arising
from the charged lepton sector is also required. Thus, GUTs may
play a crucial role and, as we will show in this letter, flipped
SU(5) turns out to be an attractive candidate.  Augmented with
suitable flavor symmetry it
enables us to predict $\te_{13}$. Moreover, it turns out that
there is direct relation between the CP asymmetry in leptogenesis
and CP violation in neutrino oscillations.


Flipped $SU(5)$ GUT \cite{DeRujula:1980qc, Derendinger:1983aj}
has several desirable features which are
conspicuously absent in standard $SU(5)$. For instance, the doublet-
triplet (DT) splitting problem is nicely realized
via the missing partner mechanism  \cite{Derendinger:1983aj},
and eventually the dimension five proton decay is naturally suppressed.
Finally, the model contains right handed neutrinos (RHN) which play an
important role in neutrino oscillations.
However, in its minimal form flipped $SU(5)$ does not shed much light on
important questions related to fermion masses and mixings. In this letter
we address this shortcoming by extending the field content of the model,
and by introducing a flavor ${\cal U}(1)$ symmetry. One of our goals is to
realize bi-large neutrino mixings responsible for the atmospheric and
solar neutrino oscillations. We also wish to understand how the quark
mass hierarchies and CKM mixings arise. The model employs a double
seesaw mechanism and we find a prediction for the third leptonic mixing
angle, namely that
$\te_{13}\simeq \sq{\fr{\De m_{\rm sol}^2}{\De
m_{\rm atm}^2}} \fr{\tan \te_{12}\tan \te_{23}}{1+\tan^2 \te_{12}}\simeq 0.07$.
Moreover, the amount of CP violation in
the neutrino oscillations can be determined from the observed
Baryon asymmetry of the Universe.
An $R$-symmetry, in combination with  ${\cal U}(1)$
also plays an essential role. Among other things,
the MSSM `matter' parity arises naturally, unwanted dimension five
baryon number violating operators, as well lepton flavor violating
processes are adequately suppressed, and the MSSM $\mu $ problem can
be nicely resolved

The `matter' sector of minimal flipped $SU(5)$ contains
$$
10_{-1}^{(i)}\equiv 10_i=(q, d^c, \nu^c)_i~,~~~~
\bar 5_{3}^{(i)}\equiv \bar 5_i=(u^c, l)_i~,~
$$
\beq
{\bf 1}_{-5}^{(i)}\equiv {\bf 1}_i=e^c_i~,
\la{matterSU5}
 \eeq
($i=1,2,3$ is family index), and the transformation properties under
$SU(5)\tm U(1)$ and field content of the  multiplets is
indicated. In the `scalar' (higgs) sector we introduce
$$
\phi (5_2)=(T_\phi, h_d)~,~~~
\bar \phi (\bar 5_{-2})=(\bar T_{\bar \phi }, h_u)~,
$$
\beq
 H (10_{-1})\hs{-0.5mm}=\hs{-0.5mm}(q_H, D^c_H, \nu^c_H)~,~~
\bar H(\ov{10}_{1})\hs{-0.5mm}=\hs{-0.5mm}
(\bar q_{\bar H}, \bar D^c_{\bar H}, \bar \nu^c_{\bar H})~,
 \la{scalSu5}
  \eeq
where the $\phi $, $\bar \phi $ pair contain the  MSSM Higgs doublets,
while $H$, $\bar H$ are used for breaking $SU(5)\tm U(1)$ to $SU(3)_c\tm SU(2)_L\tm U(1)_Y$.

Next we introduce a ${\cal U}(1)$ flavor symmetry
which distinguishes the fermion families. The ${\cal U}(1)$ breaking is
achieved by the GUT singlet superfield $X$ with  ${\cal U}(1)$ charge
$Q[X]=-1$.
We assume that the scalar component of $X$ develops a VEV such that
$\fr{\lan X\ran }{M_{\rm Pl}}\equiv \ep \simeq 0.2$, where
$M_{\rm Pl}\simeq 2.4\cdot 10^{18}$~GeV is the ultraviolet cutoff scale.
The $\ep $ will play an important role as an expansion parameter.
Knowledge of the  CKM matrix elements fixes the ${\cal
U}(1)$ charges of $10$-plets (containing quark doublets
$q_i$) as follows: $ Q[ 10_i] =(a+3, a+2, a)$.
Moreover, the up-type quark and charged lepton Yukawa hierarchies
suggest the following selection of ${\cal U}(1)$ charges:
$Q[\bar 5_i] =(b+5, b+2, b)$, $Q[{\bf 1}_i]=2a-b$, and also
 $Q[\phi ]= -2a$, $Q[\bar \phi ]=-(a+b)$.
The $a$ and $b$ are undetermined numbers for time being.

With the charge assignments given above, we have for the Yukawa couplings:
\beq
\ep^{t_{ij}}10_i10_j\phi ~,~~~~
\ep^{p_{ij}}10_i\bar 5_j\bar \phi ~,~~~~
\ep^{r_i}\bar 5_i{\bf 1}_j\phi ~,
\la{DUE}
\eeq
$$
{\rm with}~~~t_{ij}=6+(i+j-i^2-j^2)/2~,~~r_i=9+i(i-9)/2~,
$$
$$
p_{ij}=12+(i-9j-i^2+j^2)/2~.
$$
The last two  terms in (\ref{DUE})  generate up-type quark and charged
lepton masses respectively, yielding the desirable hierarchies:
$\lam_t\sim 1$, $\lam_u:\lam_c:\lam_t\sim \ep^8:\ep^4:1$ and
$\lam_e:\lam_{\mu }:\lam_{\tau }\sim \ep^5:\ep^2:1$.
The CKM mixing angles have the desirable
magnitudes ($V_{us}\sim \ep $, $V_{cb}\sim \ep^2$, $V_{ub}\sim \ep^3$).
However, the first coupling
 matrix in (\ref{DUE}) gives $\fr{m_s}{m_b}\sim \ep^4$,
$\fr{m_d}{m_b}\sim \ep^6$ which are both unacceptable. In order to
obtain satisfactory values for $m_d$ and $m_s$ the field content of the model
will be extended.
Before doing this, let us note that,
in addition to  ${\cal U}(1)$ invariance, the above couplings also display
an underlying $R$-symmetry, with
the $R$-charges of the  superfields not completely fixed.
The inclusion of the scalar sector will determine some of the charges.
Taking for $H, \bar H$ states $Q[H]=a+1/2$,  $Q[\bar H]=(a+b+1)/2$
the superpotential couplings which resolve the DT
splitting problem are $\ep \phi HH +\ep \bar \phi \bar H\bar H$.
With $|\lan H\ran |=|\lan \bar H\ran |\equiv V\simeq M_{\rm GUT}$
(in $\nu^c$ direction),
the color triplets  $T_{\phi }$ and $\bar T_{\bar \phi }$ acquire
 masses $\sim \ep M_{\rm GUT}~$ by pairing with  $D^c_H$ and $\bar D^c_{\bar
 H}$ respectively \cite{Derendinger:1983aj}.
 The MSSM doublet pair $h_u, h_d$ remain massless at this stage, as desired.
Appearance of the triplet-anti triplet pair below $M_{\rm GUT}$
helps to obtain a somewhat reduced value of the strong coupling constant
$\al_s(M_Z)\simeq 0.117$ in good agreement with experiments.

{}With the above couplings and the $R$-symmetry
transformations $\Phi_i\to e^{{\rm i}R(\Phi_i)}\Phi_i$,
$W\to e^{{\rm i}R(W)}W$ ($W$ denotes the superpotential),
we assign the $R$-charges as follows:
$R(10_i)=R(H)=\al $, $R(\bar 5_i)=\bt $,
$R({\bf 1}_i)=2\al -\bt $,
$R(\phi )=\al_{\phi }$,
$R(\bar \phi )=\al -\bt +\al_{\phi }$,
$R(\bar H)=(\al +\bt )/2$, $R(X)=0$, $R(W)=2\al +\al_{\phi }$.

To resolve the problem of the $d$ and $s$ quark masses, we introduce
the following vector-like `matter' states
 \beq
 F_{\al }(5_2)=(\bar D^c_F, L_F)_{\al }~, ~~~~
 \bar F_{\al }(\bar 5_{-2})=(D^c_{\bar F}, \bar L_{\bar F})_{\al }~,
  \la{Fs}~
  \eeq
where $\al =1, 2$ labels the two  pairs $F, \bar F$.
The couplings of these states with the $10_i$ plets
will induce  mixing of $d^c$ and
$D^c_F$ states. This allows us to improve the light down-type
quark masses. Also, the relation
$M_U=m_D^T$ (arising from the second coupling of  (\ref{DUE}))
is violated, which will be important for realizing large $\nu_{\mu
}-\nu_{\tau }$ mixing. An additional singlet scalar superfield
 $S$ also plays an important role.

The relevant couplings are given by
\beq
\bar H10\bar F \phi +
(\bar HH)^2(10F H+\bar 5\bar FH)
+SF \bar F+
\bar H10F \bar \phi ~,
 \la{Fcoup}
  \eeq
where generation indices are suppressed, and the $R$-charges are as follows:
$R(\bar F_{\al } )=-R(F_{\al })=(\al -\bt )/2$,
$R(S)=\al_{\phi }+2\al=3(3\al +\bt )/2$.
With ${\cal U}(1)$ charge assignments $Q[S]=7+3(3a+b)/2$,
$Q[F_{\al }]=-(5a+b+7/2, 5a+b+9/2)$,
$Q[\bar F_{\al }]=(a-b-7, a-b-5)/2$ and $b=-3a-4/3$,
the couplings in (\ref{Fcoup}) can be written
in family space:
$$
\begin{array}{ccc}
 & {\begin{array}{cc}
\hs{-0.8cm}\bar F_1 \hspace{0.8cm} & \hspace{0.1cm} \bar F_2
\end{array}}\\ \vspace{1mm}

\begin{array}{c}
 10_1\vs{0.1cm}\\ 10_2\vs{0.1cm} \\ 10_3
 \end{array}\!\!\!\!\!\hs{-0.2cm} &{\left(\begin{array}{ccc}

\hs{0.6mm} 1&\hs{0.3cm}\ep
\\
\vs{-0.3cm}
\\
 \hs{0.6mm}0& \hs{0.3cm}1
 \\
 \vs{-0.3cm}
 \\
 \hs{0.6mm}0\hs{0.1cm}&\hs{0.3cm}0

\end{array}\right)\fr{\bar H \phi }{M_{\rm Pl}} },~
\end{array}
\begin{array}{ccc}
 & {\begin{array}{cc}
\hs{-1.3cm}F_1 &\hs{0.1cm} F_2
\end{array}}\\ \vspace{1mm}

\begin{array}{c}
10_1\vs{0.1cm}\\ 10_2\vs{0.1cm}\\ 10_3
 \end{array}\!\!\!\!\!\hs{-0.2cm} &{\left(\begin{array}{ccc}

 \hs{0.6mm}\ep^2&\hs{0.3cm}\ep
\\
\vs{-0.3cm}
\\
 \hs{0.6mm}\ep&\hs{0.3cm}1
 \\
 \vs{-0.3cm}
\\
 \hs{0.6mm}0&\hs{0.3cm}0

\end{array}\right)\hs{-0.1cm}\l \fr{\bar HH}{M_{\rm Pl}^2}\r^{\hs{-0.1cm}2}H},
\end{array}  \!\!
$$
\beq
S\l F_1\bar F_1+\ep F_1\bar F_2+ F_2\bar F_2\r ~,
\label{10F}
\end{equation}
\begin{equation}
\begin{array}{ccc}
 & {\begin{array}{cc}
\hs{-1.5cm}\bar F_1 \hspace{1cm} & \hs{0.2cm} \bar F_2
\end{array}}\\ \vspace{1mm}

\begin{array}{c}
 \bar 5_1\vs{0.1cm}\\ \bar 5_2\vs{0.1cm} \\ \bar 5_3
 \end{array}\!\!\!\!\!\hs{-0.2cm} &{\left(\begin{array}{ccc}

\hs{0.6mm} \ep^2&\hs{0.3cm}\ep^3
\\
\vs{-0.3cm}
\\
 \hs{0.6mm}0& \hs{0.3cm}1
 \\
 \vs{-0.3cm}
 \\
 \hs{0.6mm}0&\hs{0.3cm}0

\end{array}\right)\hs{-0.1cm}
\l \fr{\bar HH}{M_{\rm Pl}^2}\r^{\hs{-0.1cm}2}H },~
\end{array}
\begin{array}{ccc}
 & {\begin{array}{cc}
\hs{-0.6cm}F_1 &\hs{0.15cm} F_2
\end{array}}\\ \vspace{1mm}

\begin{array}{c}
10_1\vs{0.1cm}\\ 10_2\vs{0.1cm}\\ 10_3
 \end{array}\!\!\!\!\!\hs{-0.2cm} &{\left(\begin{array}{ccc}

 \hs{0.6mm}\ep^2&\hs{0.3cm}\ep
\\
\vs{-0.3cm}
\\
 \hs{0.6mm}\ep &\hs{0.3cm}1
 \\
 \vs{-0.3cm}
\\
 \hs{0.6mm}0&\hs{0.3cm}0

\end{array}\right)\fr{\bar H\bar \phi }{M_{\rm Pl}} }.

\end{array}
\label{10F1}
\end{equation}
{}
Assuming that
$\lan S\ran \sim \ep_G^4M_{\rm GUT}$ ($\ep_G\equiv M_{\rm GUT}/M_{\rm Pl}$) and
substituting all appropriate VEVs,
from the first matrix couplings in (\ref{DUE}) and (\ref{10F}) with
field embeddings given in (\ref{matterSU5}), (\ref{scalSu5}),
(\ref{Fs}), the `big' $5\tm 5$ down quark mass matrix  has the form
\begin{equation}
\begin{array}{ccccc}
 & {\begin{array}{ccccc}
\hs{0.15cm}d^c_1 & \hspace{0.5cm} d^c_2 & \hspace{0.5cm}d^c_3 &
\hspace{0.5cm}D^c_{\bar F_1}
 & \hspace{0.3cm}D^c_{\bar F_2}
\end{array}}\\ \vspace{1mm}

\begin{array}{c}
q_1 \\ q_2 \\ q_3  \\\bar D^c_{F_1} \\ \bar D^c_{F_2}
 \end{array}\!\!\!\!\!\hs{-0.2cm} &{\left(\begin{array}{ccccc}

 \ep^6h_d~ & \ep^5h_d ~ &
 \ep^3h_d~ &\ep_Gh_d ~ & \ep_G\ep h_d
\\
 \ep^5h_d~ &\ep^4h_d~ &\ep^2h_d~ &
0~ &\ep_Gh_d
 \\
 \ep^3h_d~ &\ep^2h_d~ & h_d~ & 0~ & 0
 \\
 V\ep_G^4\ep^2~&V\ep_G^4\ep ~ &0~ &V\ep_G^4 ~ &V\ep_G^4\ep \\

 V\ep_G^4\ep ~ &V\ep_G^4 ~ &0~ &0~ & V\ep_G^4

\end{array}\right)}~,
\end{array}  \!\!  ~~~~
\label{flipDbig}
\end{equation}
Integrating out the
heavy $D^c_{\bar F}, \bar D^c_{F} $ states, (\ref{flipDbig})
reduces to the $3\tm 3$  matrix
\begin{equation}
\begin{array}{ccc}
 & {\begin{array}{ccc}
\hs{-6mm}d_1^c \;\;\; &\hs{0.3cm} d_2^c  \;\;\; &\hs{0.3cm} d_3^c
\end{array}}\\ \vspace{1mm}
M_d\simeq
\begin{array}{c}
q_1  \\ q_2  \\ q_3
 \end{array}\!\!\!\!\!\hs{-0.2cm} &{\left(\begin{array}{ccc}

 \ep_G\ep^2~~ &~~ \ep_G\ep  ~~  &~~\ep^3
\\
 \ep_G \ep  ~~&~~\ep_G ~~ &~~\ep^2
 \\
 \ep^3~~&~~\ep^2 ~~  &~~ 1

\end{array}\right)h_d}~,
\end{array}  \!\!  ~~~~~
\label{DL}
\end{equation}
which yields  the  desired hierarchies
$\fr{m_s}{m_b}\sim \ep_G$, $\fr{m_d}{m_b}\sim \ep_G\ep^2$.

The couplings in
(\ref{10F}) and (\ref{10F1}) yield the neutrino Dirac $5\tm 5$ matrix
\begin{equation}
\begin{array}{ccccc}
 & {\begin{array}{ccccc}
\hs{0cm}\nu^c_1 & \hspace{0.6cm} \nu^c_2
 & \hspace{0.6cm}\nu^c_3  &
\hspace{0.4cm}\bar L_{\bar F_1}
 & \hspace{0.4cm}\bar L_{\bar F_2}
\end{array}}\\ \vspace{1mm}

\begin{array}{c}
l_1 \\ l_2 \\ l_3  \\L_{F_1} \\ L_{F_2}
 \end{array}\!\!\!\!\!\hs{-0.2cm} &{\left(\begin{array}{ccccc}

 \ep^8h_u& \ep^7h_u &\ep^5h_u &V\ep_G^4\ep^2 &V\ep_G^4\ep^3
\\
 \ep^5h_u &\ep^4h_u &\ep^2h_u &0 &V\ep_G^4
 \\
 \ep^3h_u &\ep^2h_u & h_u &0  &0
 \\
 \ep_G\ep^2h_u& \ep_G\ep h_u&0 &V\ep_G^4  &V\ep_G^4\ep  \\

 \ep_G\ep h_u &\ep_Gh_u&0 &0 &V\ep_G^4

\end{array}\right)}~.
\end{array}  \!\!  ~~~~~
\label{flipML}
\end{equation}
Integrating out the heavy $L, \bar L$ states,  we obtain
\begin{equation}
\begin{array}{ccc}
 & {\begin{array}{ccc}
\hs{-0.65cm}\nu^c_1 & \hs{0.35cm}\nu^c_2  &\hs{0.35cm} \nu^c_3
\end{array}}\\ \vspace{1mm}
m_D=
\begin{array}{c}
l_1 \\ l_2 \\ l_3
 \end{array}\!\!\!\!\!\hs{-0.2cm} &{\left(\begin{array}{ccc}

 \ep_G\ep^4 & ~\ep_G\ep^3  &~\ep^5
\\
 \ep_G\ep &~\ep_G &~\ep^2
 \\
 \ep^3 &~\ep^2  &~ 1

\end{array}\right)h_u}~.
\end{array}  \!\!  ~~~~~
\label{flipmD}
\end{equation}
This modified form for the Dirac mass matrix will be important for
bi-large neutrino mixings.
With only the $\nu^c$ RHN states in (\ref{flipmD}),
one expects $\te_{23}\sim \ep^2$ (the ratio of (2,3) and (3,3) elements).
However, imagine that $\nu^c_3$ state decouples with some new singlet
state $N_3$
at high scale. Then $\te_{23}$ will be determined  by the ratio of (2,2) and
(3,2) elements which is naturally large $\sim \ep_G/\ep^2\sim 1$. By the
same token the $\nu^c_1$ state should decouple. This decoupling mechanism was
discussed in \cite{Shafi:1998dv} and successfully applied within various GUTs
\cite{Shafi:1998jf, Shafi:1999au, Shafi:2005rd}.
For large solar mixing an important role is played by the
strong mixing between the $L_2$ and $l_2$ states.
Therefore, for bi-large neutrino mixings we introduce
three additional GUT-singlet RHN $N_i$.
They will generate a suitable neutrino mass texture through the double seesaw
mechanism, after introducing the scalar superfield $\bar S$. With $R$ and
${\cal U}(1)$ charges given by
$R(N_{1,3})=3\al +\bt $, $R(N_2)=(3\al +\bt )/4$,
$R(\bar S)=3(3\al +\bt )/4$, $Q[N_1]=-17/6~,~~~Q[N_2]=-1/3$,
$Q[N_3]=1/6$, $Q[\bar S]=-3/2$,
the relevant couplings involving $N_i$ states read
$$
10_1(N_1+S\ep N_2+\ep^3N_3)\bar H+
 10_2(\bar SN_2+\ep^2N_3)\bar H +
$$
\beq
10_3N_3\bar H +(\bar HH)^2N_2N_2~,
\label{flip10N}
\end{equation}
where for higher order operators the
cut off scale $M_{\rm Pl}$ has been omitted.
The $9\tm 9$ mass matrix for neutral
fermions is given by
\begin{equation}
\begin{array}{ccc}
 & {\begin{array}{ccc}
\hs{-0.15cm}\nu \hspace{0.2cm} & \hspace{0.15cm}
\nu^c \hspace{0.2cm}
& \hspace{0.15cm}N
\end{array}}\\ \vspace{1mm}

\begin{array}{c}
\nu \\ \nu^c \\ N
 \end{array}\!\!\!\!\!\hs{-0.2cm} &{\left(\begin{array}{ccccc}

 0 & m_D &0
\\
 m_D^T&0 &M
 \\
 0&M^T & M_N
\end{array}\right)}~,
\end{array}  \!\!  ~~~~~
\label{nubigFlip}
\end{equation}
where $M$ and $M_N$ are given by $10\cdot N$ and $N\cdot N$ couplings of
eq. (\ref{flip10N}).
Integrating out the heavy $\nu^c$, $N$ states leads to the light neutrino mass
matrix given by the double seesaw formula:
$m_{\nu }^{(0)}=m_D\fr{1}{M^T}M_N\fr{1}{M}m_D^T$.
Substituting $\ep^2 \sim \ep_G$ and assuming
$\lan \bar S\ran \sim \ep_G^3M_{\rm Pl}$, we find
\beq
\begin{array}{ccc}
 & {\begin{array}{ccc}
\hs{-0.6cm} & &
\end{array}}\\ \vspace{1mm}
m_{\nu }^{(0)}\simeq
\begin{array}{c}
 \\  \\
 \end{array}\!\!\!\!\!\hs{-0.2cm} &{\left(\begin{array}{ccc}

 \ep^6 & ~\ep^3  &~\ep^3
\\
 \ep^3 &~\bt^2 &~\al \bt
 \\
 \ep^3 &~\al \bt  &~ \al^2

\end{array}\right)m}~,
\end{array}
\la{flipm0}
\end{equation}
with $m\sim \fr{(h_u^0)^2}{M_{\rm Pl}\ep_G^2}=(0.01-0.1)$~eV.
This is indeed the desired form for the leading part of the neutrino mass
matrix responsible for large
atmospheric neutrino mixing angle (provided by $\al \sim \bt $). Note that
the scale $m$ in
(\ref{flipm0}) has the correct magnitude.

{}For generating the  sub-leading part of the neutrino mass matrix,
responsible for large solar neutrino mixing, we employ the
mechanism of single RHN dominance
\cite{Suematsu:1996mk}. We
introduce an additional right handed state ${\cal N}$ and scalar
superfield $S'$ with $R$ and ${\cal U}(1)$ charges
$R[{\cal N}]=-(3\al +\bt )k/2$, $R[S']=(3\al +\bt)(k+1)/2$,
$Q[{\cal N}]=-(3+2k)/6$, $Q[S']=(22+k)/3$,
where $k$ is an integer.
The relevant couplings are
 \beq
 \ka (\bar HH)^k{\cal N}\bar 5_1\bar \phi H +
 \ka' {\cal N}F_2\bar \phi S'+
 \ep SS'(\bar HH)^{k-1}{\cal N}^2,
 \la{calNfl}
 \eeq
where $\ka $, $\ka'$ are dimensionless couplings.
We will  assume that $\lan S'\ran \sim \ep_G^{2k+1}M_{\rm Pl}$.
Recalling that the $L_2$ state (from $F_2$) strongly mixes with  $l_2$,
integrating out ${\cal N}$ gives
the sub-leading contribution to the neutrino mass matrix:
\beq
\begin{array}{cc}
& {\begin{array}{ccc} \hs{-1.2cm}~ &~  &~
\end{array}}\\ \vspace{2mm}
\begin{array}{c}
  \\ \\

\end{array} &{m_{\nu }^{(1)}=\left(\begin{array}{ccc}
\bar \al^2 &~\bar \al \bar \bt &~0
\\
\bar \al \bar \bt &~\bar \bt^{2}&~0
\\
 0 &~0 &~0
\end{array}\right)m' ~,}
\end{array}
\label{flipm1}
\end{equation}
with
$m'\sim \fr{\ka^2(h_u^0)^2}{M_{\rm Pl}\ep_G^2\ep }=
5\cdot (10^{-3}-10^{-2})$~eV (for $\ka \sim \ka' \sim 1/5$)
to explain the solar neutrino anomaly.

With the neutrino mass matrix $m_{\nu }=m_{\nu }^{(0)}+m_{\nu }^{(1)}$, with
entries $m_{\nu }^{(0)}$ and $m_{\nu }^{(1)}$ given by
(\ref{flipm0}) and (\ref{flipm1}) respectively \cite{FrIb, Asim},
the two mixing angles $\te_{12}$ and $\te_{23}$ are naturally large,
while the third leptonic mixing angle is
\beq
 \te_{13}\equiv |U_{e3}^l|\simeq \sq{\fr{\De m_{\rm sol}^2}{\De
m_{\rm atm}^2}} \fr{\tan \te_{12}\tan \te_{23}}{1+\tan^2 \te_{12}}~.
 \la{flip13pred}
 \eeq
Since the contribution to $U_{e3}^l$ from the
charged lepton sector is of order  $\ep^3$ and  can be
safely ignored, the model predicts the third leptonic mixing angle to be
$\te_{13}\simeq 0.07$.

The mass of the RHN ${\cal N}$ state
$M_{\cal N}\simeq \ep_G^{4k+4}\ep M_{\rm Pl}$ for $k=1(2)$
is of order $50$~GeV($0.5$~keV).
A keV mass sterile neutrino may contribute to
the dark energy budget of the universe \cite{Asaka:2005an}.

If the last coupling in (\ref{calNfl}) is generated by exchange of
some additional states, then $k=0$ is also possible. This gives
$M_{\cal N}\sim 10^9$~GeV, a scale  preferred by leptogenesis, and
which allows one more prediction. The lepton asymmetry is created by
the out of equilibrium decay of ${\cal N}$, in which the states
$l_i$, $\nu^c_2$ and $N_2$ are also involved. There is only one CP
violating phase in this system which also appears in the light
neutrino mass matrix, whose dominant part is generated via
$\nu^c_2$, $N_2$ states. This allows one \cite{allows} to relate the
CP asymmetry $\ep_{\cal N}$ and the leptonic Jarlskog invariant
${\cal J}^l$. For $M_{\cal N}=10^9$~GeV we have \beq
\fr{n_{B}}{s}\simeq -4.8\cdot 10^{-9}{\cal J}^l~. \la{Basym} \eeq
{}To obtain $n_B/s\simeq 9\cdot 10^{-11}$ we need ${\cal J}^l\simeq
-0.02$. CP violation of this size can be tested experimentally.

The RHNs provide a source for lepton flavor violating
rare processes such as
$l_{\al } \to l_{\bt }\ga $ \cite{Borzumati:1986qx}.
Below $M_G$, there are two right-handed
states $\nu^c_2$ and ${\cal N}$. The latter couples with the light
neutrinos so weakly ($\sim \ka \ep_G^3$ and $\ka \ep_G^5$ for $k=1$ and
$k=2$ resp.)
that it plays no role in rare processes. As far as the state
$\nu^c_2$ is concerned, its mass generated via mixing with $N_2$ is
$M_{\nu^c_2}\sim M_G\ep_G^3$.
 Assuming $N=1$ SUGRA and universality of soft scalar
masses at $M_G$, the non-universal contributions are
generated at the weak scale.
With the Yukawa couplings $\ep_Gh_u\nu^c_2(\ep^3l_1+l_2+l_3)$
one expects
\beq
{\rm BR}(\mu \to e\ga )\sim {\rm BR}(\tau \to e \ga )
\sim \ep^6{\rm BR}(\tau \to \mu \ga )~.
\la{demBr}
\eeq
{}For  $\tan \bt \sim 60$ (suggested by the charged
fermion sector), we find
${\rm BR}(\mu \to e\ga)\stackrel{<}{_\sim }5\cdot 10^{-14}$
(with sparticle masses $m_S\stackrel{>}{_\sim }100$~GeV), which is well
below the current experimental bound \cite{{Brooks:1999pu}}, but
within striking range of ongoing (and planned) experiments \cite{morbal}.
From (\ref{demBr}) the processes
$\tau \to \mu \ga $, $\tau \to e\ga$ are adequately suppressed,
consistent with the recent experimental bounds \cite{Aubert:2005ye}.

The symmetry $R\tm {\cal U}(1)$ plays another important role.
It forbids $Z_2$ 'matter' parity violating operators such as
$\bar 5\bar \phi H$, $\phi \bar F$, $\bar \phi F$,
$10\cdot 10F$, ${\bf 1}\cdot \bar 5\cdot \bar 5H$, etc, which are otherwise
allowed by flipped $SU(5)$.
Thus, 'matter' parity emerges automatically in our scheme and we have a
stable cold dark matter candidate (LSP).

We now turn to the discussion of proton decay.
Since the color triplets from scalar superfields have mass terms
$\ep M_{\rm GUT}(T_{\phi }D^c_H+\bar T_{\bar \phi }\bar D^c_{\bar H})$,
the emergence of appropriate $d=5$ $B$-violating operators require matter
couplings with $D^c_H$, $\bar D^c_{\bar H}$ fragments. However, such
couplings are strongly
suppressed ($<\hs{-2mm}\ep_G^4\hs{-2mm}\sim \hs{-2mm}10^{-8}$)
and are not relevant for nucleon decay. The Planck scale
suppressed $B$-violating operators
$\fr{\Ga_{ijkm}}{M_{\rm Pl}}
10_i10_j10_k\bar 5_m$,
$\fr{\hat{\Ga }_{ijk\al }}{M_{\rm Pl}^2}
10_i10_j10_kF_{\al }\bar H$,
$\fr{{\cal R}_{ijkm}}{M_{\rm Pl}}
10_i\bar 5_j\bar 5_k{\rm 1}_{-5}^{(m)}$ are allowed by
flipped $SU(5)$, but $R\tm {\cal U}(1)$ helps to suppress them.
One can check out that the operators
$q_1q_1q_2l_{2,3}$ and $d^ct^cu^ce^c_i$ (arising from
$\Ga $, $\hat{\Ga }$ and ${\cal R}$ resp.) are suppressed:
$\hat{\Ga }_{1122}\fr{\lan \bar H\ran }{M_{\rm Pl}}\hs{-1mm}\sim \hs{-1mm} 
\Ga_{1122}\sim \ep^2\Ga_{1123}=\ep^8\fr{\lan \bar HH\ran }{M_{\rm Pl}^2}
\stackrel{<}{_\sim }5\cdot 10^{-10}$,
${\cal R}_{131i}=\ep^7\fr{\lan \bar HH\ran }{M_{\rm Pl}^2}
\stackrel{<}{_\sim }2\cdot 10^{-9}$.
Thus, $d=5$ nucleon decay rates are adequately suppressed.

Observable proton decay arises from $d=6$ operators mediated
by $X, Y$ gauge bosons.
Because of extra triplets (coming from $\phi $, $\bar \phi $,
$H$, $\bar H$) with masses$\sim\hs{-2mm}\ep M_{\rm GUT}$, the
meeting point (scale identified with $M_{\rm GUT}$) of $SU(3)_c$ and
$SU(2)_L$ gauge couplings is reduced
by a factor $1.2$ compared to the value determined in minimal flipped $SU(5)$.
Moreover, due to the additional $F, \bar F$
states (constituting complete $SU(5)$ multiplets) the unified
coupling $\al_G$ is increased by factor $1.3$. Thus, the proton lifetime
is reduced by a factor $5$ (or so) compared to the minimal
scheme. Following \cite{Ellis:2002vk} this leads us to predict
$\tau_{p\to \pi^o e^{+}}\hs{-1mm}\simeq \hs{-1mm}10^{34-36}$~yrs.

To summarize, we have proposed an extension of
flipped $SU(5)$ which preserves the successful features of the minimal
scheme. The extension consists of vector-like 'matter'
states, and a new symmetry $R\tm {\cal U}(1)$
which insures that
matter parity is automatic, rare decay processes are adequately suppressed,
unwanted $d=5$ baryon number violation is absent,
and the MSSM $\mu $ term can be generated through one of two distinct
mechanisms \cite{Giudice:1988yz}.

The extension also enables us to reproduce observed charged fermion mass
hierarchies and the CKM mixing elements. Neutrino mass scales compatible with
present observations are also reproduced as well as bi-large mixings in the
neutrino sector.
The latter allows
to relate the cosmological CP phase  with neutrino oscillation's
CP violation (estimated to be few percent).
This prediction together with  $\te_{13}\simeq 0.07$ and
$\tau_{p\to \pi^0 e^{+}}\simeq 10^{35\pm 1}$~yrs
hopefully can be tested in the near future.

Z.T. thanks the Bartol Research Institute for warm hospitality. This
work is partially supported by DOE under contract DE-FG02-91ER40626.


\end{document}